\documentclass[
reprint,
 amsmath,amssymb,
 aps,
prb,
superscriptaddress
]{revtex4-2}

\usepackage{graphicx}
\usepackage{dcolumn}
\usepackage{bm}
\usepackage[colorlinks=true, letterpaper=ture, pdfstartview=FitV, linkcolor=blue, citecolor=blue, urlcolor=blue]{hyperref}
\usepackage{algorithmic}
\usepackage{algorithm2e}
\usepackage{listings}
\usepackage{xcolor}
\def\etal{~\textit{et~al.}}

\begin{document}

\title{Exploring the energy landscape of aluminas through machine learning interatomic potential}

\author{Lei Zhang}

\affiliation{Guangdong Provincial Key Laboratory of Magnetoelectric Physics and Devices, State Key Laboratory of Optoelectronic Materials and Technologies, Center for Neutron Science and Technology, School of Physics, Sun Yat-Sen University, Guangzhou, 510275, China}%

\author{Wenhao Luo}
\affiliation{Guangdong Provincial Key Laboratory of Magnetoelectric Physics and Devices, State Key Laboratory of Optoelectronic Materials and Technologies, Center for Neutron Science and Technology, School of Physics, Sun Yat-Sen University, Guangzhou, 510275, China}

\author{Renxi Liu}

\author{Mohan Chen}
\affiliation{HEDPS, CAPT, College of Engineering, Peking University, Beijing, 100871, China}

\affiliation{Academy for Advanced Interdisciplinary Studies, Peking University, Beijing, 90871, China}

\affiliation{AI for Science Institute, Beijing, 100080, China}

\author{Zhongbo Yan}
\affiliation{Guangdong Provincial Key Laboratory of Magnetoelectric Physics and Devices, State Key Laboratory of Optoelectronic Materials and Technologies, Center for Neutron Science and Technology, School of Physics, Sun Yat-Sen University, Guangzhou, 510275, China}

\author{Kun Cao}
\email{caok7@mail.sysu.edu.cn}
\affiliation{Guangdong Provincial Key Laboratory of Magnetoelectric Physics and Devices, State Key Laboratory of Optoelectronic Materials and Technologies, Center for Neutron Science and Technology, School of Physics, Sun Yat-Sen University, Guangzhou, 510275, China}

\date{\today}

\begin{abstract}
  Aluminum oxide (alumina, Al$_2$O$_3$) exists in various structures and has broad industrial applications. While the crystal structure of $\alpha$-Al$_2$O$_3$ is well-established, those of transitional aluminas remain highly debated. In this study, we propose a universal machine learning interatomic potential (MLIP) for aluminas, trained using the neuroevolution potential (NEP) approach. The dataset is constructed through iterative training and farthest point sampling, ensuring the generation of the most representative configurations for an exhaustive sampling of the potential energy surface. The accuracy and generality of the potential are validated through simulations under a wide range of conditions, including high temperatures and pressures. A phase diagram is presented that includes both transitional aluminas and $\alpha$-Al$_2$O$_3$ based on the NEP. We also successfully extrapolate the phase diagram of aluminas under extreme conditions ([0, 4000] K and [0, 200] GPa ranges of temperature and pressure, respectively), while maintaining high accuracy in describing their properties under more moderate conditions. Furthermore, combined with our developed structure search workflow, the NEP provides an evaluation of existing $\gamma$-Al$_2$O$_3$ structure models. The NEP developed in this work enables highly accurate dynamic simulations of various aluminas on larger scales and longer timescales, while also offering new insights into the study of transitional aluminas structures.
\end{abstract}

\maketitle

\section{introduction}

Although ``alumina" specifically refers to substances with the chemical formula Al$_2$O$_3$, it encompasses a series of polymorphic structures, including amorphous-Al$_2$O$_3$ (a-Al$_2$O$_3$), $\alpha$-Al$_2$O$_3$ (corundum), and various transitional aluminas\cite{ReviewOFalumina1998}. These materials are crucial in applications like polishing, cutting, adsorption\cite{ROUQUEROL2014393_Adsorption}, sensing, and catalysis\cite{jackson2008metal_catalysis} due to their superior mechanical properties, corrosion resistance, electrical insulation, and, in the case of transitional aluminas, unique surface activity\cite{hudson2000aluminum_book,abou2017performance_activate-alumina}. Transitional aluminas were initially discovered during the calcination of boehmite to produce aluminum oxide via the Bayer process. The final product, $\alpha$-Al$_2$O$_3$, is the most thermodynamically stable form, with other aluminas irreversibly transforming into it during calcination. However, these other forms are stable below their formation temperatures and are thus termed transitional aluminas to differentiate them from ``metastable" forms\cite{ReviewOFalumina1998}. Currently, a series of transitional aluminas, such as $\kappa$-\cite{kappa-alumina}, $\theta$-\cite{kohn1957characterization_theta}, $\delta$-\cite{jayaram1989structure_delta}, $\chi$-\cite{chi_phase-originpaper}, $\gamma$-\cite{verwey1935_gamma}, and $\eta$-\cite{zhou1991ETAandGAMMA}Al$_2$O$_3$ have been reported\cite{ReviewOFalumina1998,hudson2000aluminum_book}.

Despite extensive industrial use and active research over the past decades, the crystal structures of transitional aluminas, especially $\gamma$-Al$_2$O$_3$, remain contentious and are not definitively determined\cite{ReviewOFalumina1998, zhou1991ETAandGAMMA, ReviewofGAMMA-alumina_prins2020,ayoola2020SAED_validation,kovarik2024structural_gamma2024}. The Bayer process dehydrates bauxite to produce alumina, resulting in transitional aluminas that are highly defective, poorly crystalline, and consist of multiple coexisting phases during continuous calcination. These factors complicate structure determination through experiments alone\cite{ReviewOFalumina1998,ayoola2020SAED_validation}, and for the phase transitions between these structures, only rough temperature ranges can be identified\cite{ReviewOFalumina1998}. On the other hand, numerical simulations, including density functional theory (DFT)\cite{kohn1965self} calculations and molecular dynamics (MD) simulations, overcome some experimental limitations. However, while DFT offers high accuracy, it encounters difficulties when dealing with complex defects and large-scale systems. MD, although computationally efficient, often relies on empirical potentials that lack accuracy and reliability. These potentials, typically derived from equilibrium thermodynamic data, are inadequate for describing systems under extreme conditions. Consequently, the phase diagram of various transitional aluminas is rarely explored in numerical simulations.

In contrast, machine learning interatomic potentials (MLIPs)\cite{BehlerMilestone2007,Behler2011atom_center,behler2011neural}, which are trained on configurations that include energies, forces, and other data derived from DFT calculations, enable MD simulations to achieve near-quantum-mechanical accuracy while maintaining computational efficiency. In the past several years, significant progress has been made in MLIPs\cite{bonati2018silicon,bartok2018machine,rowe2020accurate,deringer2020general,chen2022universal_M3GNET,song2023general,batatia2023foundation_MACE}, which opens up a wide range of applications, including but not limited to structural phase transitions\cite{cheng2020evidence}, molecular structure predictions\cite{wengert2021data}, thermal transport \cite{dong2024molecular}, and phase diagrams under extreme thermodynamic conditions\cite{richard2023ab}. Research on metal and non-metal oxides\cite{ZnO_2011,titania2020,IrO2_2020,zhao2023complex_Ga2O3,Hafnia2021,hafnia2023_liushi,silica2022npj,Silicon-oxygen2024} using MLIPs continues to emerge. The versatility, high precision, and efficiency of MLIPs make them powerful tools for studying the intricate structures of transitional aluminas. However, the application of MLIPs in the study of aluminas remains relatively limited. Current research primarily focuses on using MLIPs to explore thermal transport properties of single-crystalline $\alpha$-Al$_2$O$_3$\cite{tiwari2024accurate_alumina-thermal-conduc,rodriguez2024utilizing} and structural features of a-Al$_2$O$_3$\cite{li2020effects_AlO_x,du2021predicting}. No comprehensive MLIP studies, covering a wide range of aluminas structures, have been reported.

In this work, leveraging the neuroevolution machine learning potential (NEP) approach\cite{fan2021neuroevolution_NEP_1,fan2022improving_NEP_2,fan2022GPUMD}, we develop a NEP that comprehensively covers the configuration space of aluminas, which is rigorously tested to validate its accuracy. Using nonequilibrium thermodynamic integration (NETI) method\cite{de2001single,freitas2016nonequilibrium,cajahuaringa2022non}, we calculate phase diagrams incorporating multiple transitional aluminas that are previously underexplored. Through moderate extrapolation of the NEP, we further predict phase boundaries of high-pressure alumina phases not present in the training dataset. Furthermore, by integrating NEP with our developed structure search workflow, we provide NEP-based insights into the energetically favorable configurations and the plausibility of two models of $\gamma$-Al$_2$O$_3$, namely the Smr\v{c}ok model\cite{smrvcok-Type} and the Luo model\cite{luo2021structure_i41amd}, which have been shown to match experimental results well\cite{ayoola2020SAED_validation}. This study aims to enhance future research, enriching methodologies for investigating the structures of transitional aluminas.

\section{Generalized dataset and Potential fitting}

\subsection{Generalized dataset}

Given the structural complexity of transitional aluminas, only the crystal structures of $\theta$- and $\kappa$-Al$_2$O$_3$ are relatively well understood\cite{zhou1991ETAandGAMMA,liu1991space_kappa}, and both exhibit widespread twinning phenomena\cite{ReviewOFalumina1998}. Due to the complex structural intergrowth, the exact number of necessary variants within the $\delta$-Al$_2$O$_3$ family remains unclear, although some variants of $\delta$-Al$_2$O$_3$ have been resolved thanks to advancements in experimental techniques\cite{jp50s1-s2,jp8bs1-s2}.

\begin{figure}[htbp]
  \centering
  \includegraphics[width=1\linewidth,keepaspectratio]{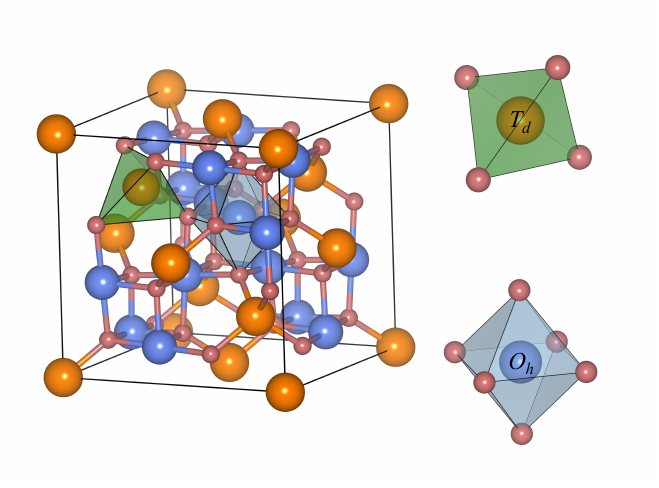}
  \caption{Schematic representation of the conventional cell of MgAl$_2$O$_4$ with the Fd$\bar{3}$m space group. The pink, orange, and blue spheres represent the O, Mg, and Al atoms located at the 32e, 8a, and 16d Wyckoff positions, respectively. The two insets on the right provide a clearer depiction of the anion coordination around the cations, where the cations at 8a are tetrahedrally coordinated ($T_d$) and those at 16d are octahedrally coordinated ($O_h$), respectively.}
\label{spinel}
\end{figure}

Verwey\cite{verwey1935_gamma} first investigated $\gamma$-Al$_2$O$_3$ in 1935, describing its structure as a traditional AB$_2$O$_4$ spinel (space group Fd$\bar{3}$m) with a lattice parameter of approximately 7.9 {\AA}. As shown in FIG. \ref{spinel}, in spinel structures such as MgAl$_2$O$_4$, oxygen atoms are arranged in a face-centered cubic sublattice (32e Wyckoff positions), with the A cations (Mg) occupying the 8a tetrahedral ($T_d$) Wyckoff positions, and the B cations (Al) occupying the 16d octahedral ($O_h$) Wyckoff positions. These 8a+16d positions are referred to as spinel sites. Summing the full-occupied Wyckoff positions in spinel model of $\gamma$-Al$_2$O$_3$ yields the stoichiometric Al$_{24}$O$_{32}$, simplified to Al$_3$O$_4$. To achieve the correct Al$_2$O$_3$ stoichiometry, 8/3 cation vacancies must be introduced per Al$_{24}$O$_{32}$, resulting in a simplified formula of Al$_{8/3}$O$_4$. This feature has led to the partial occupancy of Al cations in some $\gamma$-Al$_2$O$_3$ models. The distribution of vacancies in spinel sites has remained a controversial issue. Improved experimental studies using diffraction data or nuclear magnetic resonance indicate that vacancies are primarily at tetrahedral sites\cite{jayaram1989structure_delta,chandran2019alumina_NMR} or a mix of tetrahedral and octahedral sites\cite{lee1997distribution_mixsitedependent}, while computational studies support the concentration of vacancies at octahedral sites\cite{ealet1994electronic_sitedependent}. Pinto\etal\cite{pinto-Type}, who proposed a monoclinic configuration of $\gamma$-Al$_2$O$_3$ based on DFT from a spinel configuration, also indicated that vacancies were concentrated at octahedral sites, with lattice constants (transformed to cubic symmetry) closely matching experimental data. 

Additionally, the possibility of Al cations occupying non-spinel sites (positions other than 8a and 16d) was first proposed by Zhou and Snyder\cite{zhou1991ETAandGAMMA}. They suggested a structure with Al cations at 32e sites. Since then, more non-spinel models have emerged, including the monoclinic model proposed by Digne\etal\cite{digne-Type}, which is widely used in first-principles calculations due to its smaller number of atoms although incorrectly assumes that all Al vacancies are located on tetrahedral sites\cite{ReviewofGAMMA-alumina_prins2020}. Paglia\etal\cite{paglia2003} introduced a tetragonal model with space group I4$_1$amd,  which is the maximal subgroup of Fd$\bar{3}$m, by empirically fitting neutron diffraction data.

A recent study of Ayoola\etal\cite{ayoola2020SAED_validation}, by growing high-quality single crystals of $\gamma$-Al$_2$O$_3$, has confirmed that the non-spinel model proposed by Smr\v{c}ok\etal\cite{smrvcok-Type} in 2006 fits experimental data more accurately than the previously mentioned models. Additionally, based on Rietveld refinement of electron diffraction patterns, distinct from Paglia\etal's\cite{paglia2003}, Luo\cite{luo2021structure_i41amd} proposed a new tetragonal model with I4$_1$amd symmetry. These findings suggest that non-spinel configurations, which allow Al cations to occupy more sites, offer greater structural flexibility, thus fitting experimental data better compared to the spinel models. Furthermore, more degrees of freedom may need to be considered; for instance, Kovarik\etal\cite{kovarik2024structural_gamma2024} have recently emphasized that the Al cation vacancies alone cannot fully account for the structural complexity of $\gamma$-Al$_2$O$_3$, highlighting the role of antiphase boundaries.

In addition to $\gamma$-Al$_2$O$_3$, the existence of $\chi$-Al$_2$O$_3$ has also been noted in literature\cite{chi_phase-originpaper,ReviewOFalumina1998}, but accurate structural information for this phase remains unclear\cite{chi_phase}. Moreover, $\eta$- and $\gamma$-Al$_2$O$_3$ exhibit significant structural similarities, yet the degree of order in these phases is still debated\cite{busca2014structural_book,ReviewofGAMMA-alumina_prins2020}. Notably, no widely accepted structural model for $\eta$-Al$_2$O$_3$ currently exists. Therefore, we focus on selecting well-characterized models of transitional aluminas for our dataset, which ensures that our study comprehensively covers the most relevant and characteristic phase space of aluminas. By excluding highly debated structural models, the risk of introducing misleading information is reduced.

\begin{figure*}[htbp]
    \centering
    \includegraphics[width=1\linewidth,keepaspectratio]{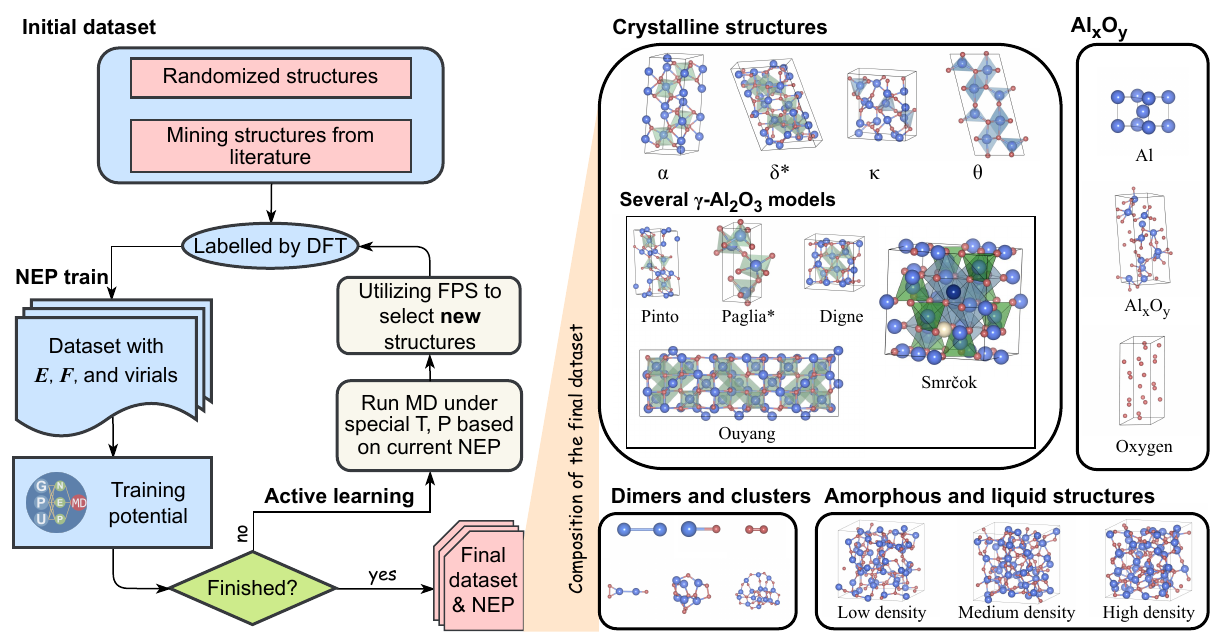}
	  \caption{A flowchart of dataset construction and potential fitting, along with the composition of the final dataset. The composition of the final dataset can be broadly categorized into four parts: crystals, non-stoichiometric Al$_x$O$_y$, dimers and clusters, and amorphous and liquid aluminas. Each part exhibits its variations. For crystals, phases or models with an asterisk (*) in the subscript indicate the existence of multiple variants, with only one variant shown in the illustration for brevity. The white sphere and deep blue sphere in the Smr\v{c}ok model represent a possible position at 16c and 48f Wyckoff positions, respectively, while the remaining atoms together form the spinel configuration. Comprehensive details of the flowchart and the final dataset are provided in the supplementary materials.}
	\label{workflowandcomposition}
\end{figure*}

As shown in FIG. \ref{workflowandcomposition}, the diagram on the right illustrates the final composition of the dataset, which can be broadly divided into four categories: crystalline phases, dimers and clusters, non-stoichiometric Al$_x$O$_y$, melt and a-Al$_2$O$_3$ with varying densities. The crystalline part of our dataset includes the $\alpha$-, $\theta$-, $\kappa$-Al$_2$O$_3$, and four variants of $\delta$-Al$_2$O$_3$ (designated as jp50s1\cite{jp50s1-s2}, jp50s2\cite{jp50s1-s2}, jp8bs1\cite{jp8bs1-s2}, and jp8bs2\cite{jp8bs1-s2} based on their respective literature sources). Additionally, we have incorporated two widely adopted $\gamma$-Al$_2$O$_3$ models as mentioned above, namely the Pinto model\cite{pinto-Type} and the Digne model\cite{digne-Type}. Furthermore, we include four $\gamma$-Al$_2$O$_3$ models with partially occupied Al cations: the Ouyang model\cite{OuyangType}, which originates from the model of Zhou and Snyder \cite{zhou1991ETAandGAMMA}, the non-spinel tetragonal model proposed by Paglia\etal\cite{paglia2003} in 2003 (referred to as Paglia2003), the local structure model proposed by Paglia\etal\cite{paglia2006fine} in 2006 (referred to as Paglia2006), and the Smr\v{c}ok model\cite{smrvcok-Type}. Additionally, although we do not consider Luo's new model with I4$_1$amd symmetry\cite{luo2021structure_i41amd}, this configuration is included when constructing structures with randomly distributed Al cations. Detailed construction specifics and the sources of all structures of various aluminas included in the dataset, along with the necessary corresponding information, are provided in the supplementary materials.

\subsection{NEP training}

\begin{figure}[htbp]
    \centering
    \includegraphics[width=1\linewidth,keepaspectratio]{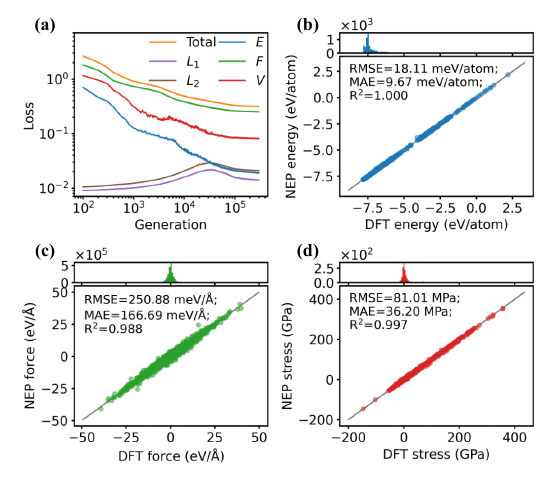}
	\caption{(a) Evolution of the various terms in the loss function for the training dataset with respect to the generation, including the $L_1$ and $L_2$ regularization, the energy ($E$) RMSE, the force ($F$) RMSE, and the virial ($V$) RMSE. (b)-(d) Reflect the fitting of NEP to the energy, force, and stress information in the dataset, where stress can be calculated by dividing the virial per atom by the volume per atom. Subfigures (b)-(d) present the regression evaluation metrics, including RMSE, Mean absolute error (MAE), and R-squared (R$^2$), with the histograms above each subfigure showing the corresponding data distribution. The solid lines in panels (b)-(d) are a guide for the eyes.}
	\label{rMSEofdataset}
\end{figure}

The NEP approach is an MLIP based on artificial neural networks (ANN), trained using the separable natural evolution strategy (SNES). According to the energy locality hypothesis proposed by Behler\etal\cite{BehlerMilestone2007}, where the energy of each atom can be determined by its local atomic environment within a cutoff radius, the NEP is trained to establish a mapping relationship between the descriptor vector with $N_\mathrm{des}$ components of the local atomic environment of a central atom $\textit{i}$ and its site energy:

\begin{equation}
	U_i\left(\{q_{\nu}^i\}_{\nu =1}^{N_\mathrm{des}}\right) = \sum_{\mu =1}^{N_\mathrm{neu}}w_{\mu}^{(1)}\tanh \left(\sum_{\nu =1}^{N_\mathrm{des}}w_{\mu\nu}^{(0)}q_{\nu}^i-b_{\mu}^{(0)}\right)-b^{(1)},
\end{equation}
where the $\tanh(*)$ represents the activation function of the hidden layer, and $N_\mathrm{neu}$ denotes the number of neurons. The parameters $w_{\mu}^{(1)}$, $w_{\mu\nu}^{(0)}$, $b^{(0)}_{\mu}$, and $b^{(1)}$ are the trainable weights and biases in the ANN of the NEP. The descriptor $q_{\nu}^i$ in NEP are composed of radial and angular components, constructed using Chebyshev polynomials and the atomic cluster expansion approach\cite{ACE_expansion}, respectively. For a more comprehensive description of the NEP approach, readers are encouraged to consult references\cite{fan2021neuroevolution_NEP_1,fan2022improving_NEP_2,fan2022GPUMD}.

The training process of the MLIPs involves minimizing a loss function, which is typically a function of the optimization target. In NEP, the loss function, $L= \lambda_e \Delta E + \lambda_f \Delta F + \lambda_v \Delta V + \lambda_1 L_1 + \lambda_2 L_2$, is designed to include the root-mean-squared error (RMSE) of the energy ($\Delta E$), force ($\Delta F$), and virial ($\Delta V$), combined with regularization terms $L_1$ and $L_2$ to prevent overfitting. The coefficients $\lambda_e$, $\lambda_f$, $\lambda_v$, $\lambda_1$, and $\lambda_2$ serve as weight factors for each term in the loss function. The flowchart of NEP training as shown on the left side of FIG. \ref{workflowandcomposition} decomposes the process into three modules based on color differentiation: initial dataset generation, NEP training, and active learning. We begin by generating structures through random perturbation and scaling of lattices at equilibrium, combined with structures mined from literature. These structures undergo first-principles calculations to establish a mapping among their energies, forces, virials, and configurations, a process referred to as ``Labelled by DFT". With these labelled structures, we fit the first-generation potential using the NEP approach and then execute the active learning process, i.e. MD simulation under specific thermodynamic conditions based on the first-generation potential, to generate new structures. Using the farthest point sampling (FPS) technique\cite{de2016comparing,bartok2017machine} based on NEP descriptor\cite{fan2022GPUMD}, the most representative new structures are pruned and selected for further DFT calculations. The newly labelled structures are added to the dataset, and the training is iteratively performed until the potential reaches the desired accuracy.

The final training dataset used for training the NEP comprises 3,335 structures and 378,762 atoms, including 2,122 crystalline structures, 423 amorphous or liquid structures, 436 non-stoichiometric Al$_x$O$_y$ structures, and 191 dimer or cluster structures. DFT calculations have been performed on all structures in the final dataset using the Vienna Ab initio Simulation Package (VASP) \cite{VASP_cite1,VASP_cite2}. The projector augmented wave (PAW) method \cite{PAW_cite1,PAW_cite2} and the PBEsol functional \cite{PBEsol_cite} were employed. Compared to the PBE functional, the PBEsol functional provides a more accurate description of self-interaction for electrons, yielding results for properties such as lattice constants and bulk modulus that are closer to experimental values. Detailed information regarding the DFT calculations and NEP training hyperparameters can be found in the supplementary materials. The training process, conducted using this dataset on GPUMD\cite{fan2022GPUMD}, is illustrated in FIG. \ref{rMSEofdataset}. FIG. \ref{rMSEofdataset}(a) shows the evolution of each parameter in the loss function during the final training. FIG. \ref{rMSEofdataset}(b)-(d) provide a more detailed depiction of the fitting accuracy and coverage range for energies, forces, and stresses, respectively. The histograms above each plot display the distribution of the corresponding data. Quantitatively, the RMSEs for the energy, force, and stress in the training dataset for this potential are 18.11 meV/atom, 250.88 meV/{\AA}, and 81.01 MPa, respectively. Results for the test dataset of different compositions are presented in the supplementary materials (FIG. S1). Overall, considering the broad range of energies, forces, and stresses covered by our NEP, these metrics demonstrate its high accuracy. We will further validate its performance.

\section{Results}
\subsection{Crystalline structures}

\begin{figure*}
	\centering
	\includegraphics[width=0.9\linewidth,keepaspectratio]{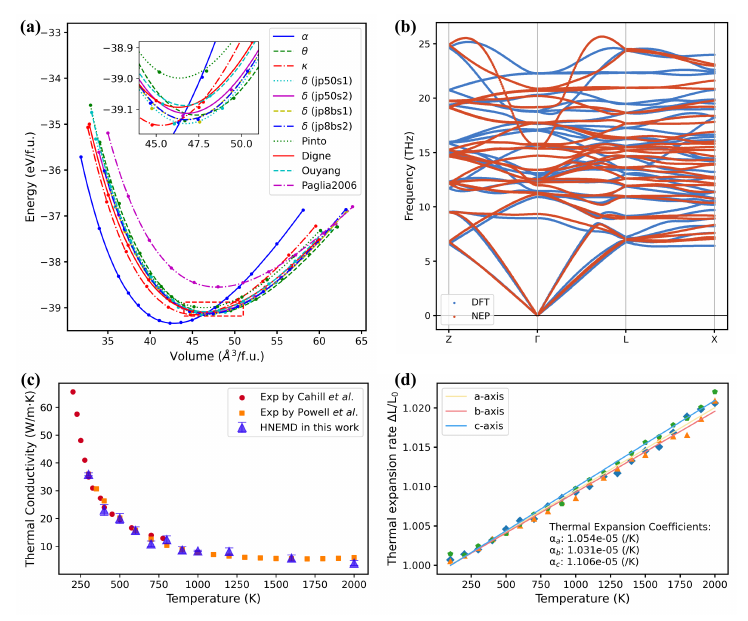}
	\caption{(a) The equation of state (EOS) curves for various alumina polymorphs were obtained by fitting the Birch-Murnaghan equation to the NEP-calculated results and were compared with the corresponding DFT-calculated points. Here, the volume and energy values are normalized per formula unit (f.u.) for comparison. The inset, which enlarges the region marked by the red dashed lines, is provided to more clearly show the energy ordering among these polymorphs. (b) Comparison of phonon spectra calculated by NEP (red) and DFT (blue). (c) The thermal conductivity of $\alpha$-Al$_2$O$_3$ was calculated using the HNEMD method at different temperatures, ranging from 300 K to 2000 K, and compared with experimental results\cite{cahill1998thermal_conductivity,powell1966thermal_conductivity}. (d) The thermal expansion coefficient of $\alpha$-Al$_2$O$_3$ was fitted during continuous temperature increases from 100 K to 2000 K. The a-, b-, and c-axis are mutually perpendicular, with the c-axis representing the [0001] crystal orientation of $\alpha$-Al$_2$O$_3$.}
	\label{crystalline_relate}
\end{figure*}

Crystalline polymorphs of aluminas, play a crucial role in the dataset and are extensively characterized, which are indispensable. In FIG. \ref{crystalline_relate}(a), the equation of state (EOS) curves of various crystalline aluminas, fitted from the NEP, are compared to the corresponding DFT calculations, serving as an important starting point for evaluating whether the potential can accurately capture the dynamics of various alumina polymorphs. The EOS curves reveal that $\alpha$-Al$_2$O$_3$ possesses the lowest equilibrium volumetric energy, followed by $\kappa$-Al$_2$O$_3$, several variants of $\delta$-Al$_2$O$_3$ and $\theta$-Al$_2$O$_3$, a few structural models of $\gamma$-Al$_2$O$_3$, and the paglia2006 model with the highest equilibrium volumetric energy. Moreover, the significant overlap of the EOS curves near the equilibrium position for these transitional aluminas reflects the possibility of phase competition or coexistence, consistent with experimental observations\cite{ReviewOFalumina1998}. Although slight deviations from DFT results remain, the NEP correctly captures the energy sequence of different structures and the main features of the EOS curves. Additionally, the elastic constants of various alumina polymorphs are tested, showing good agreement with the results from DFT calculations, as detailed in the supplementary materials (Table S2).

Phonons, as fundamental properties of materials, provide insights into thermodynamic behaviors. To verify the accuracy of the NEP in predicting thermodynamic properties of alumina, the phonon dispersions of $\alpha$-Al$_2$O$_3$ are investigated, calculated using Phonopy\cite{togo2015first_phonopy}. The phonon dispersions obtained with the NEP closely align with the DFT results in the acoustic branches. Primary discrepancies emerge in the high-frequency optical branches near the $\Gamma$ point. In ionic crystals, the macroscopic electric field induced by atomic displacements leads to the splitting of longitudinal and transverse optical (LO-TO) modes near the $\Gamma$ point. Accurately capturing the impact of this LO-TO splitting on the phonon spectra requires considering Coulomb interactions between ions through the non-analytical correction (NAC)\cite{togo2015first_phonopy,pick1970microscopic_phonon}. However, the current NEP approach does not account for charge and polarization interactions, resulting in the lack of NAC and, consequently, inconsistencies between the NEP and DFT results in the high-frequency optical branches.

Nevertheless, since the phonon dispersion relations are obtained based on the quasi-harmonic approximation and the optical branches do not dominate the thermal transport properties of most materials, the thermal conductivity of $\alpha$-Al$_2$O$_3$ is computed at various temperatures using the homogeneous nonequilibrium MD (HNEMD) method\cite{fan2019homogeneous_HNEMD}, with results showing good agreement with experiments\cite{cahill1998thermal_conductivity,powell1966thermal_conductivity}, as seen in FIG. \ref{crystalline_relate}(c). This indirectly indicates that, for $\alpha$-Al$_2$O$_3$, the exclusion of the NAC does not have a significant impact on the thermal conductivity.

As shown in FIG. \ref{crystalline_relate}(d), the thermal expansion of $\alpha$-Al$_2$O$_3$ is calculated using the NEP to further demonstrate that the NEP effectively captures higher-order interactions. The fitted results for the thermal expansion coefficient, a$_x$ $\approx$ a$_y$ = $1.05\times$ 10$^{-5}$ K$^{-1}$ and a$_z$ = $1.10 \times$ 10$^{-5}$ K$^{-1}$, show excellent agreement with experimental measurement\cite{touloukian1977thermal_expansion} and successfully capture the slight anisotropy of thermal expansion in $\alpha$-Al$_2$O$_3$.

\subsection{Amorphous and liquid aluminas}

Simulating disordered materials using DFT-based methods is challenging due to high computational demand. Employing an MLIP offers a viable alternative for conducting accurate atomic-level studies of these systems. To validate the ability of our NEP in describing disordered aluminas, we conduct a series of tests, as detailed in the section below.

\begin{figure}[htbp]
    \centering
    \includegraphics[width=1\linewidth,keepaspectratio]{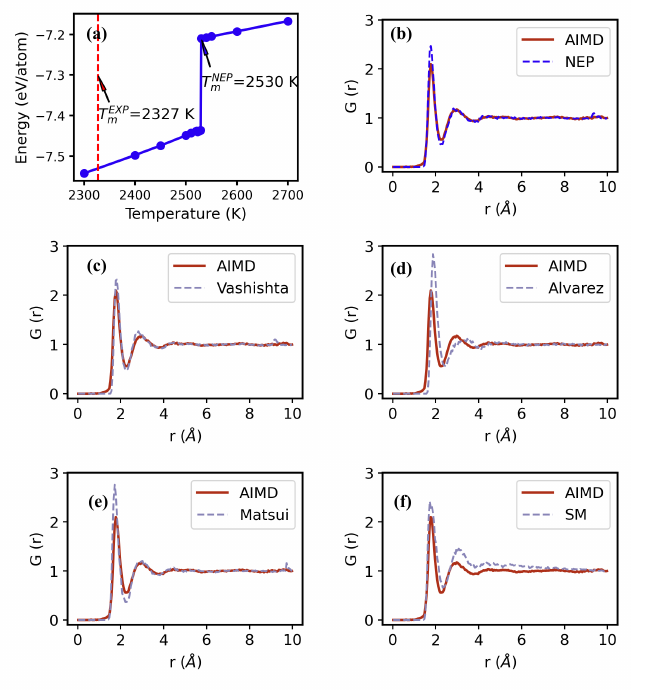}
	\caption{(a) The melting point of $\alpha$-Al$_2$O$_3$ under ambient conditions was calculated based on the two-phase method, compared with the experiment. And the comparison of RDF and AIMD results calculated by (b) NEP, (c) Vashishta, (d) Alvarez, (e) Matsui, and (f) Streitz-Mintmire (SM) for 240-atom liquid aluminas at 3000 K.}
	\label{meltandRDF}
\end{figure}

Since the process of melting $\alpha$-Al$_2$O$_3$ to obtain liquid alumina is a phase transition, we first calculat the melting point of $\alpha$-Al$_2$O$_3$ using the two-phase method. The result, 2530 K, is shown in FIG. \ref{meltandRDF}(a), which is higher than the experimentally measured value of 2327 K\cite{schneider1970cooperative_melt-point}. This discrepancy could be attributed to the fact that the crystal model used in the MD simulations is a perfect, defect-free crystal, whereas experimental samples typically contain defects, free surfaces, and interfaces in contact with the container. Taking these factors into account, the error of approximately 10\% in our NEP-based calculation is within an acceptable range.

The radial distribution function (RDF) is commonly employed to characterize short-range order in disordered systems. In this study, we utilize the RDF of liquid alumina at 3000 K, computed via \textit{ab initio} molecular dynamics (AIMD) using the PBEsol functional, as a benchmark for evaluating the accuracy of various interatomic potentials for liquid alumina. To ensure consistency in our comparisons and account for the limitations of AIMD in simulation scale, all calculations are performed on a 240-atom equilibrium supercell of $\alpha$-Al$_2$O$_3$ within the NpT ensemble (isothermal-isobaric ensemble). As illustrated in FIG. \ref{meltandRDF}(b)-(f), the AIMD-derived RDF reveals the first and second peaks, corresponding to the Al-O and O-O nearest neighbors, at 1.77 {\AA} and 3.09 {\AA}, respectively, which closely align with the experimentally reported values of 1.76 {\AA} and 3.08 {\AA} for liquid alumina\cite{ansell1997structure_liquid-alumina}, indicating that our AIMD results are highly reliable. Using the results from AIMD as a reference, we assess the empirical potentials of Vashishta\cite{vashishta2008interaction,vashishta2009erratum}, Alvarez\cite{alvarez1994surface,alvarez1995computer}, Matsui\cite{matsui1994transferable}, and Streitz-Mintmire (SM)\cite{S-M_potential} against our NEP. Notable deviations from the AIMD results are observed in the peak radii of all empirical potentials, except for the Vashishta, which, however, is slightly inferior when compared to the NEP.

When $\alpha$-Al$_2$O$_3$ melts, its structure and coordination number (CN) undergo significant changes. In $\alpha$-Al$_2$O$_3$, Al cations occupy octahedral sites, surrounded by hexagonally close-packed O anions, resulting in a high Al-O CN (n$_\textrm{AlO}$ = 6). However, in the molten disordered state, the Al-O CN deviates significantly from 6, forming a corner-sharing network dominated by 4-fold tetrahedra along with some higher-coordinated Al\cite{NEUVILLE20093410,skinner2013joint}. These structural changes also help explain the complex solidification behaviors observed when cooling the melt. For molten alumina, different quenching rates can produce various final products. According to prior studies, an a-Al$_2$O$_3$ can only form at quenching rates approaching 10$^5$ K/s\cite{levi1988phase_selection}. At relatively lower quenching rates, $\gamma$-Al$_2$O$_3$ can form directly from the melt\cite{levi1988phase_selection,ansell1997structure_liquid-alumina}, while $\alpha$-Al$_2$O$_3$ forms at quenching rates between 1 and 100 K/s\cite{weber1995solidification}. This indicates that as the quenching rate decreases, the n$_\textrm{AlO}$ of the final product increases.

\begin{figure}[htbp]
    \centering
    \includegraphics[width=1\linewidth,keepaspectratio]{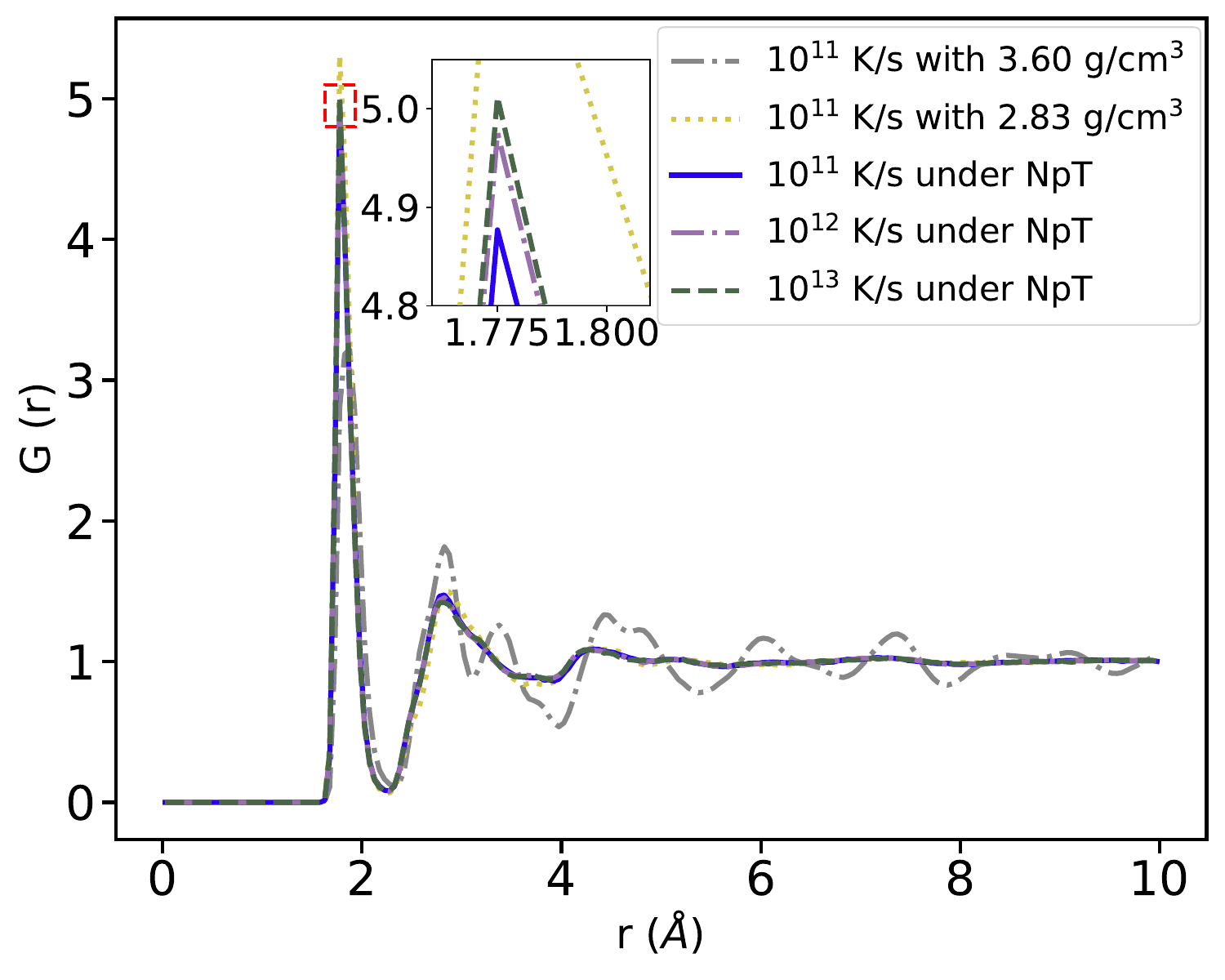}
	\caption{The influence of quenching rates and density on RDF of a-Al$_2$O$_3$. The inset, which enlarges the region marked by the red dashed lines, is reflected the relative height of the first peak of RDFs between three quenching rates. The quenching rates with densities indicate simulations conducted at fixed density, while the quenching rates under NpT indicate simulations conducted in the NpT ensemble. }
	\label{compareamorpousPartial}
\end{figure}

\begin{table}[h]
	\begin{ruledtabular}
	\centering
	\caption{Summary of average Al-O coordination number and speciation (\%) of a-Al$_2$O$_3$ with different quenching rates (K/s) or final density (g/cm$^{3}$) obtained using $R_{cut}$ = 2.2 {\AA}.}
	\label{CNs}
	\begin{tabular}{cccccc}
	Quenching rates	& 10$^{11}$ & 10$^{11}$  & 10$^{11}$  & 10$^{12}$ & 10$^{13}$\\
	\hline
	ensemble & NVT & NVT & NpT & NpT & NpT  \\
	density &3.60 &2.83 & 3.18 & 3.18 & 3.18 \\
	AlO$_4$ &25.74 &56.78 & 46.24 & 46.91 & 48.12\\
	AlO$_5$ &31.50 &39.76 & 44.26 &45.27 & 45.28 \\
	AlO$_6$ &42.73 & 3.45 & 9.49 & 7.81 & 6.59\\
	n$_\textrm{AlO}$ &5.17 & 4.47 & 4.63 & 4.60 & 4.57 \\
	\end{tabular}
\end{ruledtabular}
\end{table}

We examine the structural details of a-Al$_2$O$_3$ generated at different densities and quenching rates to assess whether the NEP can reproduce the trends observed in experiments. To ensure complete melting of high-density alumina, the melting temperature is set uniformly to 4000 K. Initially, in an NpT ensemble, $\alpha$-Al$_2$O$_3$ is held at 4000 K for 0.5 ns and then quenched to 300 K at three different rates—10$^{11}$, 10$^{12}$, and 10$^{13}$ K/s—followed by another 0.5 ns equilibration. As shown in FIG. \ref{compareamorpousPartial}, we calculate the RDF for the a-Al$_2$O$_3$ produced at the three quenching rates and find that the RDFs nearly overlap, with the only noticeable difference being that the first RDF peak, corresponding to the Al-O nearest-neighbor distance, decreases as the quenching rate slows. This suggests that slower quenching rates allow more relaxation time for the system to equilibrate, though the quenching rates used in our simulations are still much higher than those achievable in experiments due to time scale limitations. Furthermore, we find that the final density of the resulting a-Al$_2$O$_3$ is largely unaffected by the quenching rate, with all samples having a density of 3.18 g/cm$^3$, which is consistent with the experimentally reported range for a-Al$_2$O$_3$\cite{lee1995thermal, shi2019structure}. The CNs obtained for the three quenching rates presented in Table \ref{CNs} are close to those found in previous theoretical studies\cite{chandran2019alumina_NMR,shi2019structure}. Additionally, it is indicated that slower quenching rates lead to higher average CNs, which is consistent with the experimental trend mentioned earlier. The melt-quench process is studied at two different densities, 2.83 and 3.60 g/cm$^3$, using a quenching rate of 10$^{11}$ K/s. The corresponding results, as presented in FIG. \ref{compareamorpousPartial}, demonstrate that the first RDF peak for the lower density sample (2.83 g/cm$^3$) is higher and broader. Coupled with the calculated average CN of n$_\textrm{AlO}$ = 4.47, this suggests that the a-Al$_2$O$_3$ obtained from the low-density quenching resembles more closely the liquid alumina, which has a n$_\textrm{AlO}$ of 4.4\cite{ansell1997structure_liquid-alumina}. In contrast, the quenched sample with a higher density of 3.60 g/cm$^3$ exhibits features in both the RDF and CN that are more characteristic of a crystalline structure.

\subsection{Phase diagrams}

The phase transitions between $\alpha$-Al$_2$O$_3$ and its transition phases are the subject of extensive research. Most studies focus on understanding the transition pathways. The main mechanisms include diffusional and martensitic transitions\cite{ReviewOFalumina1998,bagwell2001formation}, symmetry-constrainted transition paths\cite{levin1997cubic_to_monoclinic}, and mutual transitions between the $\gamma$- and $\theta$-Al$_2$O$_3$ analyzed through first-principles calculations\cite{cai2003phase_transformation_g-to-theta}. The recently proposed ``synchro-shear" model, initially developed to describe the $\gamma$-to-$\alpha$ transition in Fe$_2$O$_3$\cite{kachi1963electron_iron-Oxides}, is adapted for explaining the $\theta$ (or $\gamma$)-to-$\alpha$ transition in alumina\cite{huang2021mechanism_synchro-shear}. However, research on the phase diagram of aluminas focuses on $\alpha$-Al$_2$O$_3$ and its potential high-pressure phases\cite{lin2004crystal_highpressure,tsuchiya2005transition_prb,caracas2005prediction,umemoto2008prediction,kato2013high}, with less attention given to the phase relationships between $\alpha$-Al$_2$O$_3$ and its transitional forms. To the best of our knowledge, only one prior theoretical study, conducted by Zhou and co-workers\cite{zhou2024first}, explores the DFT-calculated pressure-temperature (PT) phase diagram of $\alpha$-Al$_2$O$_3$ and some of its transitional phases. However, they use the quasi-harmonic approximation (QHA) to estimate phase boundaries, which may not accurately capture the dynamics of materials at high temperatures. Additionally, their study does not account for $\delta$-Al$_2$O$_3$ or other partial-occupancy structural models of $\gamma$-Al$_2$O$_3$, and lacks a detailed discussion on structural phase transitions of aluminas under high-temperature and high-pressure conditions.

\begin{figure*}[htbp]
    \centering
    \includegraphics[width=1\linewidth,keepaspectratio]{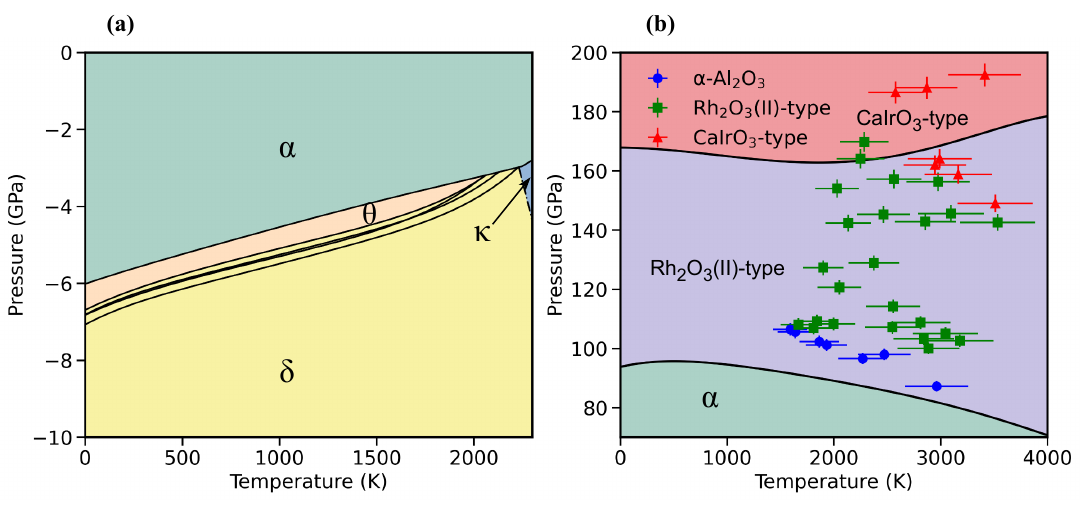}
	\caption{Phase diagram of aluminas calculated by NEP. (a) The phase diagram of $\alpha$-Al$_2$O$_3$ and transitional aluminas; the dash-dot line represents the phase boundary between $\kappa$ and jp8bs1. Regarding the transition between the $\theta$- and $\delta$-Al$_2$O$_3$, at 0 K, the lines from top to bottom represent the phase boundaries between $\theta$ and jp50s1, $\theta$ and jp50s2, $\theta$ and jp8bs2, and $\theta$ and jp8bs1, respectively. (b) The phase diagram of $\alpha$-Al$_2$O$_3$ and high-pressure phases. The scatter points with error bars represent experimental results\cite{kato2013high}.}
	\label{phase_diagram}
\end{figure*}

In this section, the NETI method \cite{de2001single,freitas2016nonequilibrium,cajahuaringa2022non}, based on the NEP, is employed as it fully accounts for anharmonic effects, thereby providing a more accurate description of phase transitions. The NETI method estimates the desired free energy difference by traversing the thermodynamic path between the target and reference systems through a time-dependent process. Unlike the equilibrium TI approach\cite{kirkwood1935statistical}, the NETI method has been shown to yield accurate results using only a limited number of relatively short non-equilibrium simulations\cite{jarzynski1997nonequilibrium}. Using this method, the phase diagrams for $\theta$-Al$_2$O$_3$, four variants of $\delta$-Al$_2$O$_3$, four models of $\gamma$-Al$_2$O$_3$ (Pinto, Digne, Paglia2006, and Ouyang), and $\alpha$-Al$_2$O$_3$ are calculated. The results are shown in FIG. \ref{phase_diagram}(a), where we calculate the PT phase diagram within the ranges of [0, 2300] K and [-10, 0] GPa. It is evident that $\alpha$-Al$_2$O$_3$ remains stable around ambient conditions, indicating that other transitional aluminas are metastable under ambient conditions from a computational standpoint.

The four variants of $\delta$-Al$_2$O$_3$ exhibit nearly identical slopes for their phase boundaries relative to the $\theta$ on the PT phase diagram, which suggests that these variants share essential lattice characteristics. This finding is consistent with experimental studies, which can only distinguish these four variants through differences in intergrowth directions and stacking sequences of structural motifs\cite{jp50s1-s2,jp8bs1-s2}. Moreover, the relatively narrow phase region between the $\theta$ and $\delta$ phase suggests that some $\delta$-Al$_2$O$_3$ variants may coexist with the $\theta$ phase under certain thermodynamic conditions\cite{ReviewOFalumina1998,jp8bs1-s2}. The $\kappa$-Al$_2$O$_3$ appears at higher temperatures and minor negative pressures, aligning with its higher density characteristic (3.98 g/cm$^3$).

Within the investigated pressure and temperature range, no stable regions are identified for the Pinto, Digne, Paglia2006, and Ouyang models of $\gamma$-Al$_2$O$_3$, indicating an absence of phase boundaries between these models and $\delta$-Al$_2$O$_3$. This absence of stable regions may be attributed to the fact that $\delta$-Al$_2$O$_3$ is generally regarded as a superstructure of $\gamma$-Al$_2$O$_3$\cite{LEVIN19973659}, a concept that has consistently been reflected in the conventions of lattice constant notation found in the related literature. In these conventions, the lattice constants of $\delta$-Al$_2$O$_3$ are often described as multiples of those of $\gamma$-Al$_2$O$_3$, from the early model proposed by Lippens\etal\cite{lippens1964study} to the four $\delta$-Al$_2$O$_3$ variants used in this study. Additionally, electron diffraction studies by Jayaram and Levi\cite{jayaram1989structure} on the $\gamma$-to-$\delta$ transition reveal that this transition is a continuous process\cite{wilson1979dehydration,ReviewOFalumina1998} marked by an increase in cation ordering while the anion arrangement remains unchanged. As a result, it is challenging to determine an accurate phase boundary between $\gamma$- and $\delta$-Al$_2$O$_3$ using free energy calculations alone.

Although MLIPs typically exhibit limited extrapolation capabilities\cite{BehlerMilestone2007}, appropriate extrapolation tests can be conducted to evaluate whether phase space of the target system has been sufficiently explored. Notably, our dataset does not explicitly include high-pressure alumina phases. To assess the extrapolation capability of our NEP, the high-pressure phase diagram of aluminas is further computed. Specifically, the phase boundaries for the transitions from $\alpha$-Al$_2$O$_3$ to the Rh$_2$O$_3$(II)-type structure and from the Rh$_2$O$_3$(II)-type to the CaIrO$_3$-type structure are calculated. While there is some discrepancy between our calculated phase boundaries and the experimental results shown in FIG. \ref{phase_diagram}(b), the trend of the $\alpha$ to Rh$_2$O$_3$(II)-type alumina transition is consistent with experimental observations. Besides, experimental studies by Funamori\etal\cite{funamori1997high} and Lin\etal\cite{lin2004crystal_highpressure} observe that the transition from $\alpha$ to the Rh$_2$O$_3$(II)-type alumina occurs near 90 GPa, which shows a better agreement with our calculated results. Additionally, the phase boundary lies within the range reported in earlier theoretical studies employing different approximations\cite{tsuchiya2005transition_prb}. At 1000 K and 1800 K, the Clapeyron slopes ($\mathrm{d}P/\mathrm{d}T$=$\Delta S/\Delta V$) for the $\alpha$-Al$_2$O$_3$ to Rh$_2$O$_3$(II)-type transition are -3.90 and -6.15 MPa/K, respectively, compared to theoretical values of -3.6\cite{tsuchiya2005transition_prb} and -6.5$\pm$1.5\cite{kato2013high} MPa/K reported in literature.

The predicted phase boundary for the Rh$_2$O$_3$(II)- to CaIrO$_3$-type alumina transition aligns well with previous calculations in the low-temperature range (0-2000 K)\cite{tsuchiya2005transition_prb}, although the accuracy diminishes above 2000 K. At 1000 K, the Clapeyron slope for the Rh$_2$O$_3$(II)- to CaIrO$_3$-type alumina transition is -4.04 MPa/K, compared to a literature value of -9.4 MPa/K\cite{tsuchiya2005transition_prb}. These results indicate that, while the former transition is accurately captured within the precision range, the latter shows a larger deviation as temperature increases. However, qualitatively, the NEP correctly predicts negative Clapeyron slopes for both phase transitions (at least up to 2000 K for the Rh$_2$O$_3$(II)- to CaIrO$_3$-type alumina transition), indicating that the trend of increasing vibrational entropy across the phase transitions is accurately captured. Importantly, despite the absence of high-pressure alumina phases in our dataset, we are able to accurately predict the phase boundaries of alumina under moderate high pressure. This suggests that our NEP is well-trained, and the structural space is comprehensively explored during the construction of the final dataset.

\subsection{Structures of $\gamma$-alumina}

Handling the inherently defective structures of $\gamma$-Al$_2$O$_3$ presents significant challenges in both experimental and theoretical studies. While structural models are developed based on physical intuition to closely fit experimental data, accurately determining the distribution of cations within these models remains difficult. The complex energy landscape arising from defect distribution further complicates the search for energetically favorable structures using conventional numerical methods. Additionally, at larger scales, new degrees of freedom, such as antiphase boundaries that influence structural stability, emerge\cite{kovarik2024structural_gamma2024}. Traditional structure search methods typically rely on constraints related to symmetry or thermodynamic conditions, rendering them unsuitable for addressing the above complexities. Furthermore, DFT methods, due to their high computational cost, are not capable of evaluating energetically favorable features in large-scale structures. However, with the advent of MLIPs, it is feasible to explore energetically favorable structures in such inherently defective systems. By using the cation distribution as a constraint, MLIPs enable an efficient and affordable search for energetically favorable structures within this complex energy landscape.

Paglia exhaustively analyzes structural models of $\gamma$-Al$_2$O$_3$ with space groups Fd$\bar{3}$m or I4$_1$amd. It is found that when cations occupy spinel sites in specific proportions, there are approximately 1.47 billion possible configurations\cite{PhysRevB.71.224115}. Clearly, when considering more occupation sites, the number of possible combinations increases dramatically, even under certain constraints. This complexity in $\gamma$-Al$_2$O$_3$ has prompted us to develop a structure search workflow based on the differential evolution algorithm\cite{storn1997differential}. In this approach, we treat the possible site distributions within the corresponding space groups of $\gamma$-Al$_2$O$_3$ as variables subject to crossover, mutation, and iteration. The NEP is then used to optimize the structures, with the resulting optimized energy serving as a criterion for structural stability.

As previously mentioned, the Smr\v{c}ok model has recently been confirmed to better align with experimental results on high-quality single-crystal samples of $\gamma$-Al$_2$O$_3$. Under this model, a structural search for the cation distribution is conducted based on the lattice parameters and other essential information provided in Ref. \cite{smrvcok-Type}. Using this workflow and leveraging the NEP, we identify energetically stable $\gamma$-Al$_2$O$_3$ structures in approximately 500 generations of a structure search with a population size of 30 per generation, although additional generations are performed to ensure convergence (see FIG. S3 in the supplementary materials). Subsequently, the 20 lowest-energy configurations are selected from the identified structures for further processing and analysis, as shown in Table \ref{smrcok_possible}.

\begin{table}[h]
	\begin{ruledtabular}
	\centering
	\caption{Summary of the distribution of cations amongst the possible site positions. The Wyckoff positions and possible numbers of sites where Al cations can occupy in the Smr\v{c}ok model with 160 atoms supercell, the maximum and minimum occupancy (occ.) numbers of these sites found in the actual structure search, and the 2$\sigma$ range were provide.}
	\label{smrcok_possible}
	\begin{tabular}{ccccc}
	Wyckoff position	& 8a & 16c  & 16d  & 48f \\
	\hline
	Possible occ. (/supercell) & 24 & 48 & 48 & 144  \\
	Min occ. &21 &0 & 37 & 0 \\
	Max occ. &24 &3 & 41 & 2 \\
	Average  &22.75 &1.25 & 39.20 &0.8  \\
	2$\sigma$ range &21.09-24.4 &-0.8-3.3&37.2-41.2&-0.8-2.4 \\
	\end{tabular}
\end{ruledtabular}
\end{table}

By comparing the occupation number of spinel sites to that of non-spinel sites, we find that the average ratio obtained from the 20 lowest-energy structures is 97:3, slightly higher than the 94:6 reported in Smr\v{c}ok's original paper\cite{smrvcok-Type}. This discrepancy might be due to the presence of high-energy occupancies in the experimental sample, where local atomic constraints prevent full relaxation. Luo\cite{luo2021structure_i41amd} proposes a structure with the I4$_1$amd space group, based on the Paglia2003 model, which is also reported to be in good agreement with experimental spectroscopic data. However, our results based on Luo model deviate significantly from the reported ratio. This difference could be due to the presence of nearest-neighbor 4a-8d cation pairs\cite{equiSite} in the Paglia2003 model, which are not allowed due to strong Coulomb repulsion between these cation pairs\cite{ReviewOFalumina1998}. In contrast, we do not observe such nearest-neighbor cation pairs in the low-energy structures generated under the Smr\v{c}ok model.

\section{Discussion}

In this work, we focus on the aluminas, developing a nearly universal MLIP based on the NEP approach. The developed NEP is subjected to rigorous testing, covering a wide range of properties from equilibrium states to extreme thermodynamic conditions. We calculate the elastic constants and fit the equations of state for all alumina polymorphs included in the final dataset, comparing the results with those obtained from DFT calculations. Taking $\alpha$-Al$_2$O$_3$ as an example, we compare and analyze the phonon spectra calculated using NEP and DFT. The accurate description of anharmonic interactions of phonons in $\alpha$-Al$_2$O$_3$ by NEP is confirmed through simulations of thermal expansion and thermal conductivity.
Using non-equilibrium thermodynamic integration method, we construct a phase diagram that includes $\alpha$-Al$_2$O$_3$ and its transitional aluminas. Furthermore, through moderate extrapolation, we obtain a phase diagram of high-pressure alumina structures that are not present in the final training dataset. The analysis of the phase diagram under high pressure fully demonstrates the robustness of our NEP.

After validating the reliability and accuracy of the potential, we make efforts to gain a deeper understanding of the structure of $\gamma$-Al$_2$O$_3$. Given its structural complexity and recognizing the limitations of existing methods in addressing such problems, we develop a search workflow that can effectively identify energetically favorable structures in systems containing intrinsic vacancies. The workflow allows us to quickly examine the correctness of the Smr\v{c}ok model and the Luo model, two models that have recently been shown to agree well with spectroscopic measurements. We provide an NEP-based understanding of the accuracy of both structural models. This workflow enables high-precision validation of existing $\gamma$-Al$_2$O$_3$ models and provides a rapid evaluation method for designing new, viable $\gamma$-Al$_2$O$_3$ structures. It can be extended to systems of any size, enabling the study of larger-scale structural features.

We hope that our work on aluminas, along with the datasets and resources we have developed, as well as the structure search workflow designed for vacancy-structured systems, will inspire further research and understanding of transitional aluminas, particularly the structure of $\gamma$-Al$_2$O$_3$. Building on the dataset we have prepared, we plan to expand our investigations to include other transitional alumina phases not discussed in this work, and to further explore the high-pressure phases of alumina, clarifying their structural details and evolutionary characteristics.

\section*{Data availability}

The final potential of aluminas and associated DFT reference data, as well as relevant structural data, are openly available in the Zenodo repository at \url{https://doi.org/10.5281/zenodo.13850687}.

\section*{Code availability}
Alongside the data, the structure search workflow we developed for this project has been made publicly available at \url{https://github.com/luowh35/DELTA}

\section*{Acknowledgements}

This research was supported by the Guangdong Basic and Applied Basic Research Foundation (Grants No. 2022A1515011618), Fundamental Research Funds for the Central Universities, Sun Yat-sen University (Grants No. 23ptpy158), Guangdong Provincial Key Laboratory of Magnetoelectric Physics and Devices (No. 2022B1212010008).

\bibliography{cite}

\begin{thebibliography}{111}
\bibitem [{equ()}]{equiSite}%
  \BibitemOpen
  \href@noop {} {}\bibinfo {note} {The equivalent spinel sites in a tetragonal lattice with I4$_1$amd symmetry are the 4a position (tetrahedrally coordinated) and the 8d position (octahedrally coordinated) in the Wyckoff notation.}\BibitemShut {Stop}%
\end{thebibliography}

\end{document}